\documentclass[%
 reprint,nofootinbib,
preprintnumbers,
 amsmath,amssymb,
 aps,
]{revtex4-1}

\usepackage{graphicx}
\usepackage{dcolumn}
\usepackage{easyReview}
\usepackage{cancel}
\usepackage{bm}
\usepackage[citecolor=green]{hyperref}
\newcommand{\SModelS}{{\tt SModelS}~2.2.1}
\newcommand{\met}{  \cancel{E}_{T} }
\newcommand{\mbl}{  m_{b\ell}}
\newcommand{\bmnames}{{\tt ON1, OFF1, OFF2, T1, T2}}

\newcommand{\gev}{\GeV}
\newcommand{\GeV}{\text{GeV}}
\providecommand{\tabularnewline}{\\}

\begin{document}

\preprint{ CERN-TH-2023-241}
\title{The rise and fall of light stops in the LHC top quark sample}
\author{Emanuele Bagnaschi$^{\S, \dagger}$, Gennaro Corcella$^{\dagger}$, Roberto Franceschini$^{\ddagger}$, Dibyashree Sengupta$^\dagger$}
\email{emanuele.angelo.bagnaschi@lnf.infn.it}
\email{\\gennaro.corcella@lnf.infn.it}
\email{roberto.franceschini@uniroma3.it}
\email{dibyashree.sengupta@lnf.infn.it}

\affiliation{%
$\S$ CERN, Theoretical Physics Department, 1211 Geneva 23, Switzerland\\
$^{\dagger}$ INFN, Laboratori Nazionali di Frascati, Via E. Fermi 40, 00044 Frascati (RM), Italy \\
$^{\ddagger}$ 
Universit\`a degli Studi Roma and INFN, Via della Vasca Navale 84, 00146, Roma
}%

\date{\today} 
\begin{abstract}
We discuss the possibility that light new physics in the top quark
sample at the LHC can be found by investigating with greater care well
known kinematic distributions, such as the invariant mass 
$\mbl$ of the $b$-jet and the charged lepton in fully leptonic $t\bar{t}$
events. We demonstrate that new physics can be probed in the rising
part of the {\it already measured} $\mbl$ distribution. To this end we analyze a concrete supersymmetric
scenario with light right-handed stop quark, chargino and neutralino.
The corresponding spectra are characterized by small mass differences,
which make them not yet excluded by current LHC searches and give rise to
a specific end-point in the shape of the $\mbl$ distribution.
We argue that this sharp feature is  general  for models of light new physics
that have so far escaped the LHC searches and can offer a precious handle
for the implementation of robust searches that exploit, rather than suffer from,  
soft bottom quarks and leptons.
Recasting public data on searches for new physics, we identify candidate models that are not yet excluded. For these models we study the $\mbl$ distribution and derive the expected signal yields, finding that there is untapped potential for discovery of new physics using the $\mbl$ distribution. 
\end{abstract}


\maketitle

\paragraph*{\bf Introduction}

Supersymmetry (SUSY) at the TeV scale is a pillar of modern Beyond Standard Model (BSM) phenomenology. 
It has shaped the thinking and the expectations for new physics for
the past few decades. After the lack of new physics signals in the
flavor experiments in the late 90s and early 2000s, great hopes were
put in the possibility to observe super-partners in high energy collisions
at the Large Hadron Collider (LHC). Indeed, it is entirely possible to build
supersymmetric models in which there is no signal of new physics in
flavor observables or in CP violation. These matters are entirely
ruled by the breaking of SUSY and there is no deep conceptual motivation
to involve SUSY in flavor dynamics. 

On the contrary, SUSY features a microscopic dynamics deeply connected
to the spontaneous breaking of the electroweak symmetry. It implies 
the existence of two Higgs doublets with fermionic partners relatively close in mass, 
thus fueling the expectation for new physics close to the mass scale of the Higgs boson.
In addition, in weakly coupled SUSY models, such as the Minimal Supersymmetric Standard Model (MSSM)~\cite{Dimopoulos:1981zb},
{the particle properties at the weak scale} are a calculable output of the microscopic properties
of the supersymmetric model~\cite{Martin:1998ve}. A prominent role
in this computation is played by the spectrum and couplings of the top
quark and its scalar partner (see, e.g., ~\cite{Slavich:2020zjv} on the role played by top and stop quarks on the MSSM Higgs masses).
Thus the investigation of new physics related to top quark phenomenology~\cite{Franceschini:2023nlp} 
emerged as a prime target for searches for new physics already at the onset of the LHC program.

Despite the great efforts put in these searches, no signs of
new physics have been spotted at the LHC yet. In light of these results,
the widespread attitude today is to favor physics scenarios with new physics characterized
by a mass scale beyond the reach of the LHC, but possibly accessible to 
future larger machines~\cite{Mangano:2651294,1812.02093,1901.10370v2,Accettura:2023aa}.
For this reason it has become customary to parameterize new physics
using contact operators than encode micro-physics in a similar fashion
to how the Fermi four-fermion contact interaction preluded to the
$SU(2)$ weak interactions~\cite{WarsawBasis:2010es}.

In spite of the general trend, in this letter we take a complementary
attitude and investigate the possibility of light new
physics in the top quark sector still not excluded at the LHC. We find that it is still possible for new physics to appear in signals that are are quite similar to the simplest manifestations that one can
imagine in the MSSM. Furthermore, we will provide new methods to probe this
enticing possibility and show that there 
is a significant potential to make a discovery in the current LHC dataset.

The majority of searches have so far
concentrated on signals characterized by \emph{large energy releases}
in the new physics events, in the TeV range, giving rise to BSM signal
that appear in regions of the phase-space where the SM has a scarce
rate. Much less activity has been devoted to searches in SM-rich signal regions. 
To fill this gap it is necessary to pursue a search method that confronts
the SM backgrounds where they are largest, that is to say in events
where the energy release is in the range of \emph{tens or few hundreds
of GeV}. In this region of phase-space, thanks to the enormous progress
in SM high-precision calculations, it is possible to carry out measurements
with exquisite precision, e.g. \cite{ATLAS:2023fsi},\cite{ATLAS:2018qzr, Andari:2017skg, ATLAS-Collaboration:2017ac},
\cite{CMS-PASFTR-18-034,CMS-Collaboration:2015xw}, \cite{ATLAS:2019onj,ATLAS-Collaboration:2017af, CMS:2016hdd}.
Therefore, in this letter we propose to carry out new searches for
BSM in regions of phase-space and in physical observables that were
previously used \emph{only} for SM measurements. Following our novel use of  the data acquired for these measurements,
we demonstrate that it is possible to obtain sensitivity to new physics
scenarios that have not been probed yet by current searches or suffer from large
uncertainty in the reach of these searches.

 {To ascertain what new physics scenarios are currently probed by the present results of the LHC, we will  recast  publicly available data using a simplified-model approach~\cite{Kraml:2013mwa,MahdiAltakach:2023bdn}. This method offers a reproducible and relatively reliable procedure to determine what new physics models, beyond those explicitly tested by the experiments, can be considered as  excluded. This recast allows us to focus our attention on the models  that are likely still experimentally allowed.  Our search for new physics models not yet excluded by recast of public information has also the value to be a stress-test of the present strategy for the publication of experimental results and their reinterpretation. We believe this is a valuable contribution to the assessment of the quality of the reinterpretation effort carried out by the LHC community.}

\paragraph*{\bf Elusive New Physics in top-quark samples}
New physics in the top quark sector has been searched for in a large number of final states (see e.g. Ref.~\cite{Franceschini:2023nlp} for a recent review). A large fraction of the signatures involves top quarks accompanied by some extra handle characteristic of new physics productions, e.g. $t\bar t+\met$ or $t\bar t+\phi$, with $\met$ being missing transverse momentum and $\phi$ a new  boson. Other possible signatures are pursued with bottom quarks, that are expected from the weak $SU(2)$ symmetry, and final states from off- and on-shell top quarks. Generally these searches are sensitive to new physics that results in large energy release, because of large mass differences between the new states themselves and between new states and SM ones. Unfortunately this search strategy hits a blindstpot when the spectrum has small mass differences.

The issue of possible blind spots in LHC searches, due to small mass differences in the new physics spectrum and possible closeness of the new physics masses with the SM ones, has emerged early on in the exploitation of the LHC data~\cite{Fan:2011cr,Fan:2015bh}. Solutions that have been suggested involved the precision measurement of the total cross-section for $t\bar{t}$ production~\cite{Eifert:2014yu,Cohen:2019ac}, possible disagreement in the extraction of $m_{t}$~\cite{Cohen:2019ac} and angular distributions \cite{Cohen:2018aa,Han:2012ve} that may be sensitive to new physics, as well as features in the kinematic distributions of very high-$p_T$ top quarks~\cite{Macaluso:2015cl}. 

Despite these proposals, the status of weak scale supersymmetry, including that of superpartners charged under QCD, remains unset.   {One reason is that not all of the above proposals have yet been incorporated in the suite of searches that the experiments carry out, which can be interpreted in a twofold manner. On the one hand this is signaling there is still room for improvement, on the other hand this can be read as a signal of the objective difficulty to use new observables for searches, e.g. for the difficulty to obtain the desired data with sufficiently small uncertainty as required by the proposed new searches. This motivates our effort to seek a method that requires mere reinterpretation of measurements already carried out and no new observables to be measured with high accuracy.}

 {To fill in this gap we propose a new method to identify new physics hidden in the top quark sample. Our method leverages the notable feature of the candidate new physics models to involve energy releases in the decay of the new physics particles that are typically smaller than those of the SM $t\bar{t}$ production and other SM background processes.}

\begin{figure}
    \centering
    \includegraphics[width=0.95\linewidth]{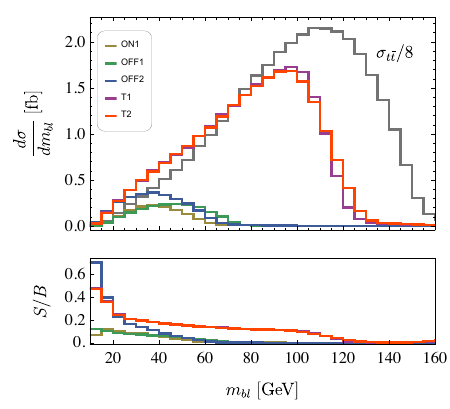}       
    \caption{The $\mbl$ distribution of the SM (grey line, rescaled by 1/8) and  MSSM signals (colored lines) for the benchmark points \bmnames.The signal distribution has a characteristic rise-and-fall shape, which makes it easier to observe. In the bottom panel we display the ratio of the signals over the SM contribution.}
    \label{fig:rise-and-fall}
\end{figure}

 {As a consequence of this feature the range of the Lorentz invariants that can be used to characterize the kinematics of the events involving new physics is different from that of the SM processes. In particular the maximum of the invariant mass of the bottom and the lepton that arise from the top quark decay, denoted by $\mbl$, turns out to be significantly smaller than in SM $t\bar{t}$ production for models that have not been excluded from present searches. This gives rise to a notable rise-and-fall shape of the new physics signal in the low energy part of the $\mbl$ spectrum. This rise and fall shape changes for each new physics model spectrum, but it is generic of the whole class of new physics models not presently excluded. Some examples of $\mbl$ spectra that arise for MSSM parameter choice not presently excluded by recast of public data are presented in Fig.~\ref{fig:rise-and-fall} for illustration.} 

 {Remarkably, the quantity $\mbl$ is routinely measured by the LHC experiments, thus our observation can be readily applied to measurements {\it already} carried out, with no required new measurements to be performed to search for new physics.} \footnote{We stress that in principle other quantities can be used to test new physics following the same logic that lead us to $\mbl$. E.g. the spectrum of the energy of the $b$ jets is sensitive to new physics~\cite{Agashe:2012bn,Franceschini:2017eyj}, but there is no published data on this quantity except a preliminary CMS measurement~\cite{CMS:2015jwa}.}

\begin{figure}
    \centering
        \includegraphics[width=0.75\linewidth]{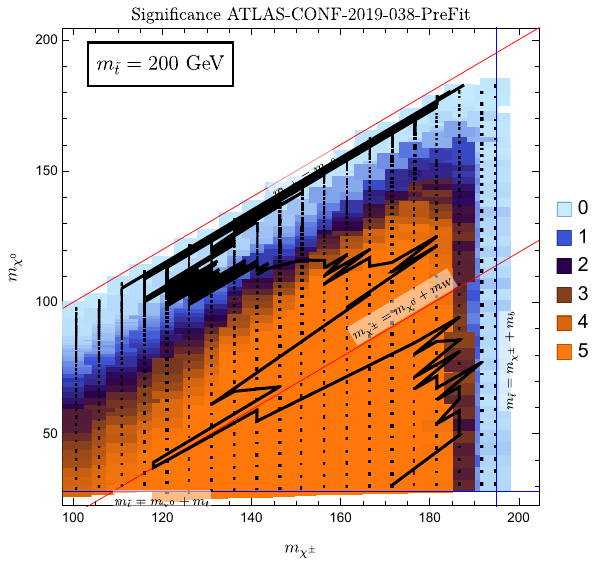}
    \caption{MSSM signal significance eq.~(\ref{eq:significance}) versus  ${\chi}_1^{\pm}$ and ${\chi}_1^{0}$ at fixed $m_{\tilde{t}_1}= 200$~GeV evaluated from the $\mbl$ spectrum and its uncertainty in ATLAS~\cite{ATLAS:2019onj} (before the fit of SM backgrounds). The black dots correspond to the points in the plane we have explicitly generated. The colored shades are step-wise interpolations of the generated points. The color scale is saturated at $z=5$. The black line is the approximate contour for the exclusion from present searches. The full result of our recast is presented in the Appendix in Fig.~\ref{fig:r-200}.  The region of the plane at the left of the black line is deemed as excluded from our recast.}
    \label{fig:significance-200}
\end{figure}

 {Based on the $\mbl$ spectrum of each new physics model we derive an estimate for the sensitivity to new physics of precision $\mbl$ measurements carried out at the LHC. We take ATLAS~\cite{ATLAS:2019onj} and CMS~\cite{CMS:2018fks} recent results on the $\mbl$ spectrum (and its uncertainty in each bin) to compute the expected statistical significance that a new physics signal from the MSSM would have. The obtained expected significance is given in Fig.~\ref{fig:significance-200} from which we can see that our method has the potential to probe large areas of the MSSM parameter space. In particular it is sensitive to mass spectra that are not currently probed by reinterpretation of LHC searches. In addition the sensitivity of our method depends on a different combination of the new physics masses with respect to standard searches. Thus our method can probe new physics in previously unexplored corners of the models space and also provide independent information on models that can be studied with already proposed strategies. This is particularly useful in view of the complexity to set bounds with traditional methods due to systematics and modelling difficulties, e.g. related to the transition from on- to off-shell intermediate resonances.}

\paragraph*{\bf Methodology}
The top quark is pair-produced at the LHC through strong interactions with a cross-section about 1~nb \cite{Czakon:2013goa,Catani:2019hip}, yielding a sample of ${\cal O}(10^{8})$ $t\bar t$ pairs produced so far. Given such large numbers, it is possible to study the top quark in great detail, including  rare decay modes. We consider the decay to a double lepton final state, which leads to a final state containing 2 $b$-jets, 2 oppositely charged leptons and their corresponding neutral partners which manifest as missing transverse momentum at the LHC. In this sample we carefully study a well-known observable, $\mbl$, the invariant mass of the lepton and the $b$-jet. This quantity
has been used to extract the mass of the top quark~\cite{CMS:2014cza, CMS:2009kup} and its width \cite{ATLAS:2017vgz,CMS:2014mxl}, demonstrating the great control that the experiments can achieve on this observable.

As argued above, the observable $\mbl$ can show deviations from the SM prediction if it gets contaminated by a new particle which has mass close to $m_t$ and can mimic the final state of fully leptonic $t\bar t$ events. Referring to the MSSM as a case study, such final states are given by events with pair-production of the lightest stop quark $\tilde{t_1}$, decaying into the lightest chargino ${\chi_1^{\pm}}$ and a $b$-quark, followed by ${\chi_1^{\pm}}$ decay into the lightest neutralino ${\chi_1^{0}}$ and lepton pair via a real or a virtual $W$:
$$
\tilde{t_{1}} \to \chi^{\pm}_{1} b \to  b \ell \nu \chi_{1}^{0}\;.
$$
 The lightest neutralino and the neutrino manifest
themselves as missing transverse momentum, so that this signals leads to the same
final state as fully leptonic $t\bar t$ pairs.
An analogous final state could be achieved in the MSSM if the lightest stop quark $\tilde{t_1}$ decays into a (leptonically decaying) $t$ quark and ${\chi_1^{0}}$. In this case the dominant decay  is ($\tilde{t_1} \to t {\chi_1^{0}}$), which suffers much stronger bounds from specific searches~\cite{CMS:2019wav} and therefore it is not in the focus of our work. 

To root our study in a concrete and reproducible setup, we consider a large set of points in the MSSM parameter space and study the signal in the $\mbl$ spectrum for each MSSM point. For concreteness we consider three possible values for the lightest stop mass: $m_{\tilde{t_1}}$ = $180$, $200$ and $220$~GeV. For each value of $m_{\tilde{t_1}}$ we scan the parameter space to get different values of $m_{{\chi_1^{\pm}}}$ and  $m_{{\chi_1^{0}}}$. In order to obtain our reference points,
we make use of {\tt SPheno}~4.0.3~\cite{Porod:2011nf}, interfaced with {\tt SARAH}~4.15.1~\cite{Staub:2013tta}. The {\tt SPheno} input parameters
are set at a high scale $Q$ and then run down to the weak scale by means
of Renormalization Group Equations. The description of the inputs used is provided in the Appendix.

Each of these points has been checked against searches available for recast using \SModelS~\cite{Alguero:2021dig}. In particular we check the value of the metric $r$ computed by \SModelS. Points for which $r>1$ are deemed to be excluded by the recast, while for $r<1$ we consider the  present public data to be insufficient to exclude that model. Clearly, it is possible that the full dataset held by the experimental collaborations, as well as combinations of signal regions not taken into account by \SModelS, can still exclude the points were we find $r<1$.

Next we simulate the contribution to $\mbl$ for each  parameter space point using {\tt Pythia}~8.3 ~\cite{Bierlich:2022pfr} in the region of phase space identified by the following selection:
\begin{eqnarray}
 & p_T(\ell) \geq 25~\GeV,& |\eta(\ell)| < 2.5, \nonumber \\
 & p_T(j) \geq 25~\GeV,& |\eta(j)| < 2.5, \label{eq:cuts}
\end{eqnarray}
for jets made with anti-kT~\cite{Cacciari:2008gp} algorithm with $R=0.4$ and  separations between jets and leptons $\Delta R(\ell, j)>0.2$, $\Delta R(j,j)>0.4$ and $\Delta R(\ell,\ell)>0.1$. This is a selection closely following that of the experimental collaborations, e.g. \cite{ATLAS:2019onj, ATLAS:2017vgz, CMS:2016hdd}, except for minor differences in the selection for $\ell=e$ and $\ell=\mu$ that we do not pursue.
We have considered variations of the cuts and found that, if attainable, softer selections on the transverse momenta would magnify the signal in the $\mbl$ distribution even further, but we limit ourselves to the conservative choice of  cuts as in eq.~(\ref{eq:cuts}).
 {The $\mbl$ spectra that we obtain are compared with the spectrum measured by ATLAS~\cite{ATLAS:2019onj} and CMS~\cite{CMS:2018fks} for 139~fb$^{-1}$ integrated luminosity. As the experimental results are endowed with an uncertainty on each bin of the measured differential cross-section $d\sigma/d\mbl$, we can use the expected rate of MSSM signal to compute  a significance
\begin{equation}
\label{eq:significance}
z = \sqrt{ \sum_{i}\left( \frac{S_i}{\delta B_i}\right)^2}\,,
\end{equation}
where $S_i$ is the MSSM signal  yield expected in the $i$-th bin of the published histogram and $\delta B_i$ is the uncertainty on each bin as published by the experiments. In absence of more precise information from the experiments, the uncertainty in each bin is assumed to be uncorrelated with the others.}

 {We note that both experimental collaborations provide two set of uncertainties: one is obtained with nominal Monte Carlo predictions and  uncertainties, while a second one is provided after the measured $\mbl$ spectrum is used as a constrain on the sum of the Monte Carlo predictions for several SM processes contributing to the relevant region of phase-space. These two results are indicated by the experiments as ``pre-fit'' and ``post-fit'' measurements of the $\mbl$ distribution. The post-fit one has smaller uncertainties and leads to stronger bounds on new physics. For reference we note that the smallest uncertainty in a single bin for the ``pre-fit'' ATLAS result we use is about 5\%. Using the ``post-fit'' result would give even stronger exclusions, as the smallest uncertainty in a single bin would be reduced to 0.8\% in that scenario. However, we argue that it should be used with care, because it is obtained assuming that the $\mbl$ spectrum is due solely to the SM and no new physics.}

 {In Fig.~\ref{fig:significance-200} we show the more conservative ``pre-fit'' result of the significance eq.~(\ref{eq:significance}) from the ATLAS result~\cite{ATLAS:2019onj}.   Points for which $z>2$ can be excluded at 95\% confidence level with the new proposed analysis of $\mbl$.} Strikingly, the region excluded by our proposal covers a large area of the chargino-neutralino mass plane not excluded by the recast of the present searches.

 {We observe that the contours of $z$ in the chargino-neutralino mass plane closely follow the contours of the maximal $\mbl$ value that can be obtained for a cascade decay \cite{Franceschini:2022vck,Barr:2010zj}, thus they depend on a different combination of the masses compared to the present searches. This is apparent comparing the contours of $r$ in Fig.~\ref{fig:r-200} in the Appendix and the contours of $z$ in Fig.~\ref{fig:significance-200}.}
 
For greater detail, in Tab.~\ref{tab:benchmarks} we present the result for several points that are not excluded by recast of present searches, i.e. \SModelS~gives $r<1$~\footnote{The most recent version of {\tt SModelS} at the time of writing is 2.3.2. We checked that the new searches included in the newest release of {\tt SModelS} do not change the values of $r$ for the points in this table.}. We note that in several cases one expects  deviations from the SM in the $\mbl$ distribution much larger than the uncertainties published by the experiments. These include cases for chargino-neutralino mass differences close to $m_W$, where the present searches have a marked blindspot. We note that CMS results tend to give a weaker sensitivity: this is due to the coarser binning of the data published by CMS with respect to ATLAS.
The table presents results for   three masses $m_{\tilde{t}}$ considered.  {The complementarity of the proposed search using $\mbl$ is evident for all the masses $m_{\tilde{t}}$ considered, as to testify the general validity of the point that we make in this letter.}
\begin{table}
    \centering
\begin{tabular}{c|cccccccc}
\ {BM} & $\mu$ &  $M_1$& $A_t$  & $m_{\chi^{+}}$ & $m_{\chi^{0}}$ & $z$~\cite{CMS:2018fks} & $z$~\cite{ATLAS:2019onj} & $r$\tabularnewline
\hline 
\hline
\hline
& \multicolumn{8}{c}{ $m_{\tilde{t}}=200$~GeV}  
     \\
\hline
\hline
 \ {ON1} & 185 & 95 & 2820.5 & 186.6 & 85.6 &
   \ {[0.8,1.7]} & \ {[2.7,14.3]} & 0.9 \\
 \ {OFF1} & 155 & 160 & 2857.5 & 156.4 & 123.3 &
   \ {[0.9,1.8]} & \ {[2.6,14.8]} & 0.7 \\
 \ {OFF2} & 175 & 145 & 2839.5 & 176.6 & 123.5 &
   \ {[1.5,3.]} & \ {[5.1,25.5]} & 0.8 \\
 \ {T1} & 135 & 65 & 2895.5 & 136.2 & 54. &
   \ {[4.,7.7]} & \ {[10.7,61.3]} & 0.8 \\
 \ {T2} & 135 & 60 & 2895.5 & 136.2 & 49.9 &
   \ {[4.1,7.9]} & \ {[10.8,60.6]} & 0.8 \\
\hline
\hline
& \multicolumn{8}{c}{ $m_{\tilde{t}}=220$~GeV}  
     \\
\hline
\hline
 \ {OFF3} & 155 & 150 & 3140.5 & 156.4 & 118.6 & \ {[0.7,1.4]} & \ {[1.9,10.9]} & 0.8 \\
 \ {OFF4} & 170 & 160 & 3122 & 171.5 & 130.8 & \ {[0.9,1.8]} & \ {[2.5,13.7]} & 0.6 \\
 \ {ON2} & 190 & 95 & 3104 & 191.7 & 86.1 & \ {[2.1,4.3]} & \ {[6.1,32.8]} & 0.7 \\
 \ {OFF5} & 190 & 145 & 3104 & 191.7 & 127.7 & \ {[1.4,2.8]} & \ {[4.2,22.5]} & 0.6 \\
 \ {ON3} & 190 & 65 & 3104 & 191.7 & 58.9 & \ {[1.9,3.7]} & \ {[5.3,28.7]} & 0.8 \\
\hline
\hline
& \multicolumn{8}{c}{ $m_{\tilde{t}}=180$~GeV}  
     \\
\hline
\hline
  \ {OFF6} & 165 & 115 & 2570.5 & 166.5 & 99.2 & \ {[1.2,2.5]} &
   \ {[4.8,22.9]} & 0.8 \\
 \ {OFF7} & 160 & 105 & 2580 & 161.5 & 90.4 & \ {[2.2,4.5]} &
   \ {[7.2,36.3]} & 0.8 \\
 \ {OFF8} & 160 & 170 & 2570 & 161.5 & 130.3 & \ {[0.6,1.2]} &
   \ {[2.4,11.2]} & 0.6 \\
 \ {OFF9} & 155 & 150 & 2579.5 & 156.4 & 118.5 & \ {[1.6,3.2]} &
   \ {[5.3,27.2]} & 0.8 \\
 \ {OFF10} & 145 & 175 & 2598.5 & 146.3 & 122.2 & \ {[0.8,1.6]} &
   \ {[2.4,12.7]} & 0.8
\end{tabular}

    \caption{Chargino and neutralino masses, input parameters $\mu$, $M_1$ and $A_t$, all given in GeV  for few benchmarks (BM). Resulting value of $r$ computed from \SModelS\, and the range of the significance eq.~(\ref{eq:significance}) expected from the $\mbl$ spectrum analysis using ATLAS~\cite{ATLAS:2019onj}  or CMS~\cite{CMS:2018fks} measurements. The low (high) end the significance range corresponds to uncertainties on the $\mbl$ spectrum before(after) a fit using SM predictions for the known backgrounds.}
    \label{tab:benchmarks}
\end{table}

\paragraph*{\bf Summary and outlook}
 The presently available public information from LHC searches for new physics can be re-interpreted to test models for which the LHC experiments have not provided explicit results.
The MSSM is the model for which the majority of the efforts to provide reinterpretation data has been carried out so far.
In our work we have used the publicly available information and assembled it to the extent that it is possible with a standard tool such as \SModelS. We have investigated the bounds on weak scale supersymmetry spectra featuring light $SU(2)$ singlet stops squarks and light bino/Higgsino states with masses close to the top quark mass. We find that the LHC search results, reinterpreted by using the most recent data, still cannot exclude large part of the models we tested. 

At face value this result implies that the LHC experiments, even after years of efforts,
have not yet reached a level at which weak-scale supersymmetry can be said to be ruled completely, not even in the case of colored super partners. 

 Exclusion of supersymmetric models with light stops, bino and Higgsino may be possible using the full power of the data that is held by the experimental collaborations, but this cannot be deduced neither from publicly available information on searches results nor from a recast of the public material. Therefore, we urge the experimental collaborations to present data for these models in a similar fashion to our results in Fig.\ref{fig:r-200}, as to provide results for these specific types of models. We also advise that they should do all that is in their power to release and maintain public information suitable to obtain exclusions on new physics models in a systematic and reproducible way.
We have explained that the generation of spectra of this type in the MSSM may require a very focused setting of MSSM spectrum generators, thus a specific effort, beyond the ``wide-net'' studies in the context of the pMSSM, may be needed to tackle this issue. 

In addition to raising a flag about the actual reach of the LHC searches for new physics and their reinterpretability, we have provided an example of search strategy that would cover the gaps that we have highlighted. The novel strategy leverages the precision in top quark measurements and in particular the spectrum of the b-jet+lepton invariant mass $\mbl$, that is used to measure  the top quark mass and its width with high precision. Relying on publicly available data on the measured $\mbl$ spectrum we have identified the sensitivity contours in the chargino-neutralino mass plane at various representative values of the stop mass. Remarkably the sensitivity of our search strategy depends on a different combination of the masses that that relevant for the other searches. In particularly it seems to cope well with the transition between on- and off-shell intermediate resonances, such as $\chi^{\pm}\to \chi^{0}W^{\pm} \longmapsto   \chi^{\pm}\to \chi^{0}\ell^{\pm}\nu$. Therefore, the novel search strategy that we propose offers a very valuable complementary constraint that can fill the gaps and extend the reach of   the searches for weak scale supersymmetry that have been devised so far. Furthermore, our proposal holds the promise to become an exemplary  incarnation of the paradigm of new physics searches through high precision Standard Model measurements. Thus we urge the experimental collaborations to start interpreting the measurement of $\mbl$ and possibly other top quark precision properties as searches for new physics and to provide explicit results for bounds on models from these observables.

\section*{Acknowledgements }
The work of RF is supported in part by the MUR PRIN2022 Grant n.202289JEW4.
It is a pleasure for RF to thanks Federico Meloni, Javier Montejo Berlingen, Krzysztof Rolbiecki  for useful discussions. We thank the Galileo Galilei Institute for Theoretical Physics for the hospitality and the INFN for partial support during the completion of this work.

\bibliography{reference}
\appendix 
\section{Sensitivity of present searches}
\begin{figure}
    \centering
    \includegraphics[width=0.75\linewidth]{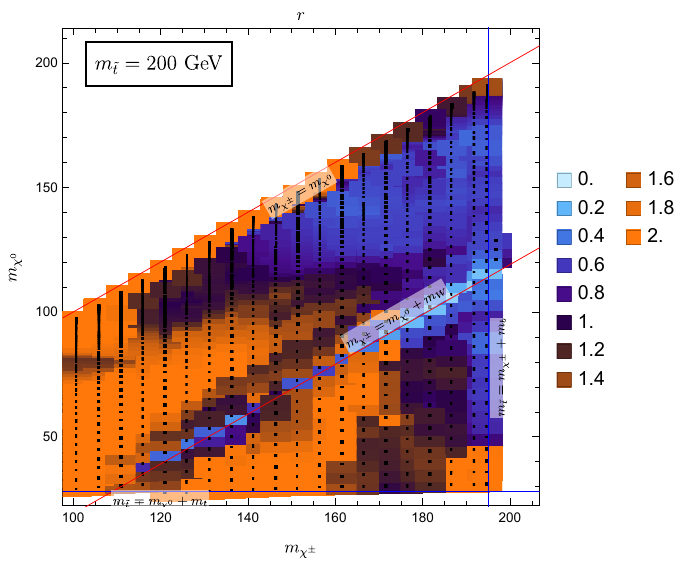}       
    \caption{Values of $r$ for different masses of $\tilde{\chi}_1^{\pm}$ and $\tilde{\chi}_1^{0}$ at fixed $m_{\tilde{t}_1}=$ 200 GeV calculated using \SModelS. A value of $r>1$ implies the model is excluded. The color scale is saturated at $r=2$.} 
    \label{fig:r-200}
\end{figure}

The full result on the exclusion metric $r$ computed by \SModelS\; is given in Fig.~\ref{fig:r-200}. 
A notable drop of the strength of the bounds from the searches appears along the line diving the on-shell and off-shell intermediate $W$ in the $\tilde{\chi}^+$ decay. Similar plots can be obtained for other values of $m_{\tilde{t}_1}$. The overall look is quite similar, thus we do not display these figures here for brevity. The general trend is for lighter $m_{\tilde{t}_1}$ larger parts of the chargino-neutralino mass plane are constrained. As clear from Tab.~\ref{tab:benchmarks} there are regions unconstrained by the recast of the present searches for all the three masses $m_{\tilde{t}_1}$ we have tested. It should be recalled that the approach of \SModelS does not allow to combine exclusions from different signal regions, as this requires to account for correlations, that are in general not available. Hence it is possible that a more complete analysis, that can only be carried out by the experiments, may find that larger regions of the chargino-neutralino mass plane are excluded. 

\section{MSSM points generation}
In general we focused our generation on spectra in which the decay via chargino is prominent, as to pursue this elusive decay mode. To obtain viable spectra from a full MSSM generator we had to set inputs at a particularly low scale. This is an indication of the fact that the models we are studying are far from being simplest options in an MSSM with a single mass scale. Clearly, if the MSSM realizes this type of spectrum, a non-trivial pattern of SUSY breaking giving rise to multi-scale super-partners spectrum must be arranged. We do not interpret this fact and we limit ourselves to use the MSSM to generate mass spectra, whose existence can be read as a mere example and proof of consistency of the new physics we are studying.  For the reproducibility of our work we list all the relevant inputs of the spectrum generator.
At $Q=1.6$~TeV, we set: 
\\
\begin{itemize}
    \item $m_{\tilde{u}}(1,1)^2$ =  $m_{\tilde{u}}(2,2)^2$ =  $m_{\tilde{q}}(i,i)^2$ =  $m_{\tilde{l}}(i,i)^2$ = $m_{\tilde{e}}(i,i)^2$ =  $m_{\tilde{d}}(i,i)^2$ = 1.2 $\cdot$ $10^7$ GeV$^2$ for $i=1,2,3$,  where $\tilde{q},\tilde{l}$ are charged under SU(2), while $\tilde{d},\tilde{u},\tilde{e}$ are charged only under hyper-charge.
    All the off-diagonal squark and slepton mass terms are set to zero;
    \item $m_{\tilde{u}}(3,3)^2$ = 1.7 $\cdot$ $10^5$ GeV$^2$ which governs the lightest stop eigenstate and results in  a $\tilde{t}_{1}$ almost pure SU(2) singlet state;
    \item $M_1\in $ [5,1000]~GeV, $M_2=1$~TeV, $M_3=3.5$~TeV;
    \item $M_A^2$ = 2 $\cdot$ $10^6$ GeV$^2$, $\tan \beta$ = 10;
    \item $\mu\in [100~\gev, m_{\tilde{t}_1}]$;
    \item $A(3,3)$: the trilinear scalar soft SUSY breaking interaction for the stop in the range $X_t + \mu\cdot\cot\beta+ [-100,100]\,\text{GeV}$  where the exact value is obtained  by trial-and-error to get the desired $m_{\tilde{t}_1}$. All other trilinear couplings are set to zero.
\end{itemize}
In order to optimize our effort we did not consider $M_1$ below 5 GeV as it tends to make the decay channel $\tilde{t_1} \to t {\chi_1^{0}}$ very copious and the spectrum is typically excluded by dedicated searches, e.g. Ref.~\cite{CMS:2019wav}. 
The lower limit on $\mu$ takes into account limits from LEP-II experiments~\cite{LEP:2003aa}.

\end{document}